# Static and Dynamic Magnetic Properties of FeMn/Pt Multilayers

Ziyan Luo[1], Yumeng Yang[1], Yanjun Xu[1,2], Mengzhen Zhang[3], Baoxi Xu[2], Jingsheng Chen[3], and Yihong Wu[1,*]

[1]*Department of Electrical and Computer Engineering, National University of Singapore, 4 Engineering Drive 3, Singapore 117583, Singapore*

[2]*Data Storage Institute, A*STAR (Agency for Science, Technology and Research), 2 Fusionopolis Way, 08-01 Innovis, Singapore 138634, Singapore*

[3]*Department of Materials Science and Engineering, National University of Singapore, 117575*

Recently we have demonstrated the presence of spin-orbit toque in FeMn/Pt multilayers which, in combination with the anisotropy field, is able to rotate its magnetization consecutively from 0º to 360º without any external field. Here, we report on an investigation of static and dynamic magnetic properties of FeMn/Pt multilayers using combined techniques of magnetometry, ferromagnetic resonance, inverse spin Hall effect and spin Hall magnetoresistance measurements. The FeMn/Pt multilayer was found to exhibit ferromagnetic properties, and its temperature dependence of saturation magnetization can be fitted well using a phenomenological model by including a finite distribution in Curie temperature due to subtle thickness variations across the multilayer samples. The non-uniformity in static magnetic properties is also manifested in the ferromagnetic resonance spectra, which typically exhibit a broad resonance peak. A damping parameter of around 0.106 is derived from the frequency dependence of ferromagnetic resonance linewidth, which is comparable to the reported values for other types of Pt-based multilayers. Clear inverse spin Hall signals and spin Hall magnetoresistance have been observed in all samples below the Curie temperature, which corroborate the strong spin-orbit torque effect observed previously.

[*] Author to whom correspondence should be addressed: elewuyh@nus.edu.sg

## I. INTRODUCTION

Multilayer structures consisting of ultrathin nonmagnetic (NM) layers, particularly Pt and Pd, and ferromagnetic (FM) layers such as Co and Fe, have been of both fundamental and technological interest since late 1980's.[1] When the thicknesses of both NM and FM layers are controlled within a certain range, typically less than 1.5 nm, the multilayer as a whole exhibits ferromagnetic properties with dominantly perpendicular magnetic anisotropy (PMA). Some of these multilayer films have already been applied in magneto-optic recording[2] and more recently also in magnetic tunnel junctions as part of the reference layer.[3,4] Stimulated by earlier work on proximity effect at the FeMn and Pt interface,[5] we have recently carried out a systematic study of FeMn/Pt multilayers.[6,7] Despite the fact that FeMn is an antiferromagnet (AFM), FeMn/Pt multilayers with ultrathin FeMn and Pt layers (< 1 nm) were found to exhibit global FM ordering with in-plane magnetic anisotropy. A large field-like spin-orbit torque (SOT) was found to be present in the multilayer when a charge current flows through it.[7] Quantification of the SOT strength was carried out by varying the thicknesses of both FeMn and Pt systematically and the results corroborate the spin Hall effect (SHE) scenario, *i.e.*, spin current is generated and absorbed by the multilayer, thereby generating the SOT. We have further demonstrated that the SOT is able to rotate the magnetization of FeMn/Pt multilayers by 360º without any external field. These results demonstrate clearly the potential of FeMn/Pt multilayers in memory and sensor applications.

In order to gain further insights into the SOT generation mechanism in FeMn/Pt multilayers, in this paper, we report on ferromagnetic resonance (FMR), inverse spin Hall effect (ISHE) and spin Hall magnetoresistance (SMR) studies of multilayer samples which exhibit clear SOT effect. Before proceeding to dynamics studies, the static magnetic properties of the multilayers were characterized using magnetometry at variable temperatures. Special emphasis was placed on the understanding of the temperature dependence of the saturation magnetization. From fitting of the experimental data using different models, it is found that the multilayers exhibit the characteristic of three-dimensional Heisenberg universality class with a finite Curie temperature distribution. This correlates well with the large linewidth of resonance peaks observed in both FMR and ISHE. A large damping parameter (~ 0.106) is derived from the frequency dependence of the FMR, which is comparable to the values reported previously for other types of Pt-based multilayers. The observation of both ISHE and SMR suggests the presence of spin current generation/absorption processes, corroborating the strong SOT effect observed previously. The role of asymmetric FeMn/Pt and Pt/FeMn interfaces in generating the SOT is discussed for samples with relatively thick FeMn and Pt layers, whereas for samples with ultrathin Pt as well as co-sputtered samples, extrinsic SHE/ISHE may play a more important role.

**II. EXPRIMENTAL**

[FeMn($t_{FeMn}$)/Pt($t_{Pt}$)]$_n$ multilayers (here $n$ denotes the repeating period) with different FeMn and Pt thicknesses, $t_{FeMn}$ and $t_{Pt}$, were prepared on SiO$_2$/Si substrates by magnetron sputtering with a base pressure of $2 \times 10^{-8}$ Torr and working pressure of $3 \times 10^{-3}$ Torr, respectively. The nominal composition of Fe:Mn is 50:50. The structural properties of the multilayers were characterized using both X-ray diffraction (XRD) and X-ray reflectivity (XRR) analysis. Magnetic measurements were carried out using a Quantum Design vibrating sample magnetometer (VSM) with the samples cut into a size of 2 mm × 2 mm. The FMR measurements were performed at room temperature via a coplanar waveguide (CPW), designed to have an impedance of 50 Ω within a broad frequency range up to 20 GHz. The waveguide, 5 mm long, has a signal line of 150 μm and a signal to ground line spacing of 20 μm. The two signal lines of the CPW were connected to a Vector Network Analyzer (VNA) via high-frequency probes. The FMR spectra were obtained by placing a 2 mm × 2 mm sample directly on the CPW with sample surface facing down and taking readings of the $S_{21}$ signal while sweeping a DC magnetic field in the signal line direction. For ISHE measurements, the samples were patterned into Hall bars with a lateral dimension of 2000 μm × 120 μm by combined techniques of photolithography, sputtering deposition and lift-off. Following the Hall bar fabrication, a 100 nm SiO$_2$ insulating layer was deposited to isolate electrical conduction between the waveguide and the multilayer with the contacts to the Hall bar uncovered for subsequent electrical measurements. The last step was to deposit a 150-μm wide and 200-nm thick Cu coplanar waveguide and four 500 μm × 500 μm contact pads. The same Hall bar was used to measure the SMR, which was obtained by rotating the samples under a constant field of 3 kOe in the *xy*, *yx*, and *zx* planes, respectively.

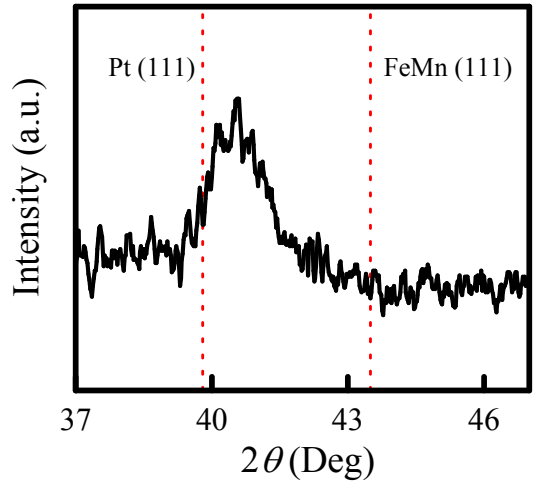

FIG. 1. X-ray diffraction pattern of Pt(1)/[FeMn(0.6)/Pt(0.4)]$_{10}$. Dotted lines indicate the (111) peak position of Pt and FeMn, respectively.

## III.    RESULTS AND DISCUSSION

### A. Structural properties

The as-deposited multilayers were characterized using both high-angle XRD and small-angle XRR. Figure 1 shows the XRD pattern of Pt(1)/[FeMn(0.6)/Pt(0.4)]$_{10}$, covering the range of bulk fcc Pt (111) peak at 39.8° and bulk fcc FeMn (111) peak at 43.5°, using the Cu Kα radiation (λ = 1.541 Å). Here the number and symbols inside the parentheses denote the thickness of individual FeMn and Pt layers in *nm*. In order to prevent oxidation, all the samples except stated otherwise were all covered by a 1 nm Pt capping layer. As can be seen from the figure, the diffraction pattern is dominated by a main peak at 40.5° – 40.6°, which falls between the bulk Pt (111) and FeMn (111) peaks. This suggests that the multilayer is (111) textured and its lattice spacing is the average of those of Pt and FeMn, though it is more dominantly of Pt characteristic. The FeMn (111) peak is almost at the same level of the baseline, which is presumably caused by the combined effect of ultrathin thickness, interface mixing and small scattering cross sections of Fe and Mn as compared to Pt. Similar phenomena have also been observed in Co/Pt multilayers, in which the peak position is near that of Pt and increases with increasing the Co thickness.[8-10] The small-angle XRR was measured with an incident angle in the range of 0° – 10° with a step of 0.02°. Figure 2 shows the XRR of a multilayer with structure: Pt(1)/[FeMn(0.6)/Pt(0.6)]$_{30}$ and another co-sputtered sample, *i.e.*, Pt and FeMn were deposited simultaneously using the same deposition time and power. The $n = 1$ Bragg maximum corresponding to a period of 1.06 nm (about 20% smaller than the nominal values) is clearly observed in the spectrum for the multilayer sample (red solid-line). In contrast, only thickness induced fringes are observed in the spectrum for the co-sputtered sample (blue dotted-line). The result demonstrates that the multilayer has a well-defined periodicity.

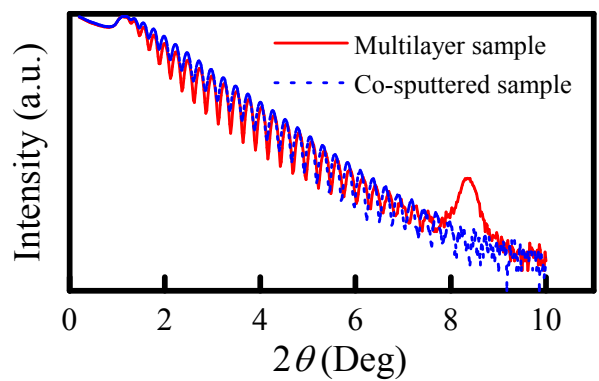

FIG. 2. XRR patterns of Pt(1)/[FeMn(0.6)/Pt(0.6)]$_{30}$ multilayer sample (red solid-line) and co-sputtered sample (blue dotted-line) deposited under the same condition.

**B. Magnetic properties**

All the multilayers with $t_{FeMn}$ < 0.8 nm and $t_{Pt}$ < 1 nm exhibit ferromagnetic properties with in-plane magnetic anisotropy. The Curie temperature ($T_C$) varies from 250 K to 380 K, depending on both the total and individual layer thicknesses. Figure 3a shows the hysteresis loop of Pt(1)/[FeMn(0.6)/Pt(0.3)]$_{10}$ at 50 K and 300 K, respectively. The coercivity at 50 K is around 240 Oe, but it decreases rapidly to about 1 Oe at 300 K. Such kind of behavior is typical of samples exhibiting FM properties above room temperature (RT). Figure 3b shows the saturation magnetization as a function of the temperature (*M-T*) with the FeMn layer thickness ($t_{FeMn}$) fixed at 0.6 nm and Pt layer thickness ($t_{Pt}$) ranging from 0.1 nm to 0.8 nm. As it was found that a minimum repeating period of 3 – 4 is required for most samples to exhibit ferromagnetic properties above RT, we fixed the repetition period for all the samples at 10. Although the polarized Pt also contributes to the measured magnetic moment, it is difficult to quantify it for samples with different thickness combinations and at different temperature. Therefore, as an approximation, we only take the overall FeMn volume into consideration when calculating the saturation magnetization. As shown in the figure, the $M_s$ at low-temperature increases with increasing $t_{Pt}$, though the sample with $t_{Pt}$ = 0.1 nm has a significantly smaller magnetization. An opposite trend is observed for $T_C$ which decreases with $t_{Pt}$, saturating at about 300 K when the adjacent FeMn layers are completely separated magnetically by the Pt layer. Both trends are in qualitatively agreement with findings reported for Co/Pt multilayers,[11] which can be accounted for by the proximity effect at Pt/FeMn interfaces. Pt is known to be just under the Stoner limit that can be readily polarized when it is in direct contact with ferromagnetic materials. In the present case, although FeMn is an AFM in bulk phase, it shall behave like a "superpara-AFM" when it is ultrathin, *i.e.*, $t_{FeMn}$ < 1 nm. This can be inferred from exchange bias studies in FeMn-based AFM/FM bilayers, which have revealed that a minimum thickness of 4 – 5 nm is required for FeMn to establish a measurable exchange bias to the FM at RT.[12] Despite its superpara-AFM nature, when it forms a multilayer with Pt, the mutual interaction at their interfaces promotes FM order in both layers which eventually extends throughout the multilayer when both layers are ultrathin. Therefore, as long as Pt is thin enough to allow complete polarization by the adjacent FeMn layers the average magnetic moment at low temperature will increase with the Pt thickness. On the other hand, the decrease of $T_C$ at increasing Pt thickness is presumably due to weakening of exchange coupling throughout the multilayer caused by incomplete polarization of the Pt layers at central regions. The anomaly at $t_{Pt} = 0.1$ nm can be readily understood by taking into account the effect of interface roughness. At this thickness, Pt is probably partially discontinuous, resulting in direct coupling of neighboring FeMn layers at certain locations and thereby reduces the saturation magnetization and $T_C$.

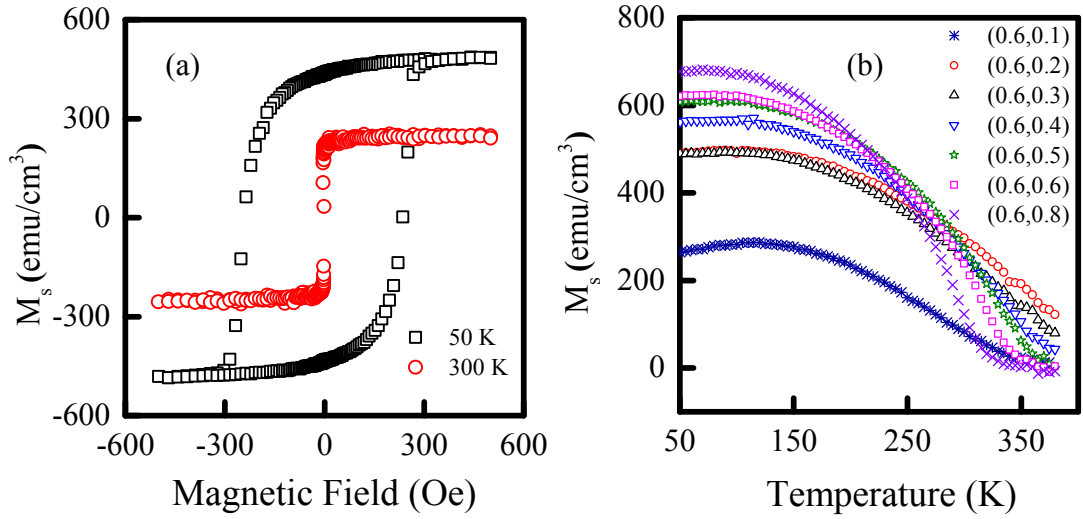

FIG. 3. (a) Hysteresis loop of Pt(1)/[FeMn(0.6)/Pt(0.3)]$_{10}$ measured at 50 K (square) and 300 K (circle), respectively. (b) Saturation magnetization as a function of temperature. The legend ($t_1, t_2$) denotes a multilayer with a FeMn thickness of $t_1$ and Pt thickness of $t_2$. The number of period for all samples is fixed at 10.

In order to gain more insights on the magnetic properties of the multilayers, we examine the *M-T* curves using different models. The temperature-dependence of magnetization for a ferromagnet at low-temperature can be calculated from the number of thermally excited magnons – quanta of spin-wave. Associated with each magnon is a magnetic moment $g\mu_B$, and therefore the total moment of magnon is given by

$$N = g\mu_B \sum_k \frac{1}{\exp\left(\frac{\hbar\omega_k}{k_B T}\right) - 1} \tag{1}$$

where $g$ is the electron g-factor, $\mu_B$ is the Bohr magneton, $\hbar$ is the reduced Planck's constant, $\omega_k$ is the magnon frequency, and $k_B$ is the Boltzmann's constant. Under the long wavelength limit, the magnon dispersion relation may in general be written as $\hbar\omega_k = Dk^n$, where $D$ is the spin-wave stiffness, and $n = 2$ for a ferromagnet and $n = 1$ for an AFM. Substitute the dispersion relation into Eq. (1), one obtains

$$\begin{aligned} N &= \frac{4\pi g\mu_B}{(2\pi)^3} \int_0^\infty \frac{k^2 dk}{\exp(Dk^n/k_B T) - 1} \\ &= \frac{1}{2\pi^2 n} g\mu_B \zeta\left(\frac{3}{n}\right) \Gamma\left(\frac{3}{n}\right) \left(\frac{k_B T}{D}\right)^{3/n} \end{aligned} \tag{2}$$

where $\zeta$ is the Riemann zeta function and $\Gamma$ is the Gamma function. Eq. (2) can be used to calculate the temperature dependence of magnetization in FM or stagger order parameter in AFM. Since the FeMn/Pt multilayers exhibit ferromagnetic properties despite the fact that bulk FeMn is an AFM, in what follows we only focus on FM. By substituting $n = 2$ into Eq. (2), we obtain the Bloch $T^{3/2}$ law, i.e.,

$$M(T) = M(0)(1 - B_{3/2} T^{3/2}) \tag{3}$$

where $B_{3/2}$ is a constant proportional to $D^{-3/2}$. Although the Bloch $T^{3/2}$ law can satisfactorily explain the M-T dependence at low temperature, it fails at high temperature because of the neglect of magnon-magnon interactions and deviation of the dispersion relation from $\hbar\omega_k = Dk^2$ at large $k$. For a Heisenberg ferromagnet, the high-temperature effect may be included in $M(T)$ by introducing a temperature-dependent $D$, namely, $D(T) = D(0)(1 - B_{5/2} T^{5/2})$, where $B_{5/2}$ is a constant.[13] As a result, the $M(T)$ in a wide temperature range can be modelled by

$$M(T) = M(0)\left[1 - B_{3/2}\left(\frac{T}{1 - B_{5/2} T^{5/2}}\right)^{3/2}\right]. \tag{4}$$

When $B_{5/2}$ is small, $M(T)$ can be approximated as

$$M(T) = M(0)\left(1 - B_{3/2}T^{3/2} - \frac{3}{2}B_{3/2}B_{5/2}T^4\right) \tag{5}$$

Although Eq. (5) improves the fitting at higher temperature as compared to the Bloch $T^{3/2}$ law, it is still unable to reproduce the *M-T* curve in the entire temperature range, and the deviation from experimental value tends to increase near $T_C$ due to the critical behavior of ferromagnet.

In order to improve the fitting near $T_C$ by taking into account the critical behavior, we invoke the semi-empirical model developed by M. D. Kuz`min,[14] which turned out to be very successful in fitting the *M-T* curves of many different types of magnetic materials. According to this model, the temperature-dependent magnetization of a ferromagnet is given by:

$$M(T) = M(0)\left[1 - s\left(\frac{T}{T_C}\right)^{3/2} - (1-s)\left(\frac{T}{T_C}\right)^{5/2}\right]^{\beta} \tag{6}$$

where *M(0)* is the magnetization at zero temperature, $T_C$ is the Curie temperature, *s* is the so-called shape parameter with a value in the range of 0 – 2.5, and $\beta$ is the critical exponent whose value is determined by the universality class of the material: 0.125 for two-dimensional Ising, 0.325 for three-dimensional (3D) Ising, 0.346 for 3D XY, 0.365 for 3D Heisenberg, and 0.5 for mean-field theory[15,16]. On the other hand, for surface magnetism, $\beta$ is in the range of 0.75 – 0.89.[17,18] The shape parameter *s* is determined by the dependence of exchange interaction, including its sign, on interatomic distance in 3D Heisenberg magnets.[19] This may have implications to multilayer samples as lattice distortion and strain are unavoidable at the interfaces due to large lattice match between and FeMn and Pt.

The *M-T* dependence shown in Fig. 3b can be fitted reasonably well using Eq. (6) with $\beta$ = 1.01 ~ 2.55 and *s* = -0.85 ~ -0.45, except that the fitted magnetization drops to zero more quickly as compared to the experimental data. The large $\beta$ values seem to suggest that the *M-T* of FeMn/Pt multilayers follows the surface scaling behavior. However, a careful examination of the results suggests that this may not be the case because we found that $\beta$ decreases as $t_{Pt}$ increases. An opposite trend would have been observed should it were due to surface mechanism because a thick Pt layer would help enhance the 2D nature of ferromagnetism at the interfaces. This prompted us to consider other possible factors that may affect the shape of *M-T* of the multilayers in a more prominent way as compared to the case of a uniform 3D ferromagnet. The one that came into our attention is the high sensitivity of $T_C$ to the Pt thickness as manifested in the M-T curves in Fig. 3b; this may lead to finite distribution of $T_C$ throughout the multilayer due to thickness variation induced by interface roughness. When this happens, the magnetization may drop more slowly near $T_C$, as observed experimentally. To this end, we modified Eq. (6) by including a normal distribution of $T_C$, which leads to

$$M(T) = M(0) \int_0^\infty \left[ 1 - s \left( \frac{T}{T_C} \right)^{3/2} - (1-s) \left( \frac{T}{T_C} \right)^{5/2} \right]^\beta \frac{1}{\sqrt{2\pi}\Delta T_C} \exp\left[ -\frac{(T_C - T_{C0})^2}{2\Delta T_C^2} \right] dT_C \qquad (7)$$

where $T_{C0}$ is the mean value of $T_C$ and $\Delta T_C$ is its standard deviation. As shown in Fig.4a, all the *M-T* curves can be fitted very well using Eq. (7) with a fixed $\beta$ value of 0.365, especially near the $T_C$ region. Note that $\beta = 0.365$ is the critical exponent for 3D Heisenberg ferromagnet. For the sake of clarity, all the curves in Fig. 4a except for the one for $t_{Pt} = 0.1$ nm are shifted vertically. In the figure, symbols are the experimental data and solid-lines are fitting results. The fitting values for *M(0)*, $T_{C0}$, and $\Delta T_C$, and *s* as a function of Pt thickness are shown in Fig. 4b, 4c, and 4d respectively. Except for the sample with smallest $t_{Pt}$, the trends of *M(0)* - $t_{Pt}$ and $T_C$ - $t_{Pt}$ are opposite with each other, *i.e.*, the former increases whereas the latter decreases with $t_{Pt}$. Both are manifestation of the fact that the global FM ordering in FeMn/Pt multilayers originates from the proximity effect at FeMn/Pt interfaces, as discussed above. It is interesting to note that $\Delta T_C$ also increases when $t_{Pt}$ decreases, and importantly, the range of $\Delta T_C$ for samples with $t_{Pt} = 0.1 - 0.8$ nm corresponds to the range of average $T_C$ of all samples with $t_{Pt}$ ranging from 0.1 nm to 0.8 nm. These results are consistent with the $T_C$ fluctuation scenario, *i.e.*, a larger fluctuation in $T_C$ is expected in samples with smaller $t_{Pt}$ due to interface roughness and its range should be corresponding to the difference in average $T_C$ when $t_{Pt}$ varies from 0.1 to 0.8 nm or less. Another important result derived from curve fitting is the $t_{Pt}$

- dependence of the shape parameter *s*. According to M. D. Kuz`min *et al.*, for 3D Heisenberg magnets, *s* is determined by the dependence of exchange interaction on interatomic distance.[19] It is generally positive with a small *s* (< 0.4) corresponding to metallic FMs with long-range ferromagnetic ordering and high $T_C$, whereas a large *s* (> 0.8) is indicative of competing exchange interactions and the resultant material typically has a low $T_C$. As shown in Fig. 4d, *s* is small and positive for samples with $t_{Pt}$ = 0.6 nm and 0.8 nm, but it turns negative for smaller $t_{Pt}$. When *s* is negative, the $T^{3/2}$ term of the base of Eq. (6) becomes positive, or in other words, it contributes positively to *M(T)* when temperature increases. This is counterintuitive for 3D Heisenberg ferromagnet. It suggests that, in addition to isotropic exchange coupling, interfacial Dzyaloshinskii-Moriya interaction (DMI) may play a role, particularly in samples with smaller $t_{Pt}$. As DMI favors non-collinear alignment of spins, a weakening of DMI at moderately elevated temperature may give a relative boost of isotropic exchange coupling, thereby resulting in a positive contribution to the magnetic moment at intermediate temperature range. This may explain why *s* is negative, though further studies are required to quantify the effect of DMI on temperature dependence of magnetization in these multilayers.

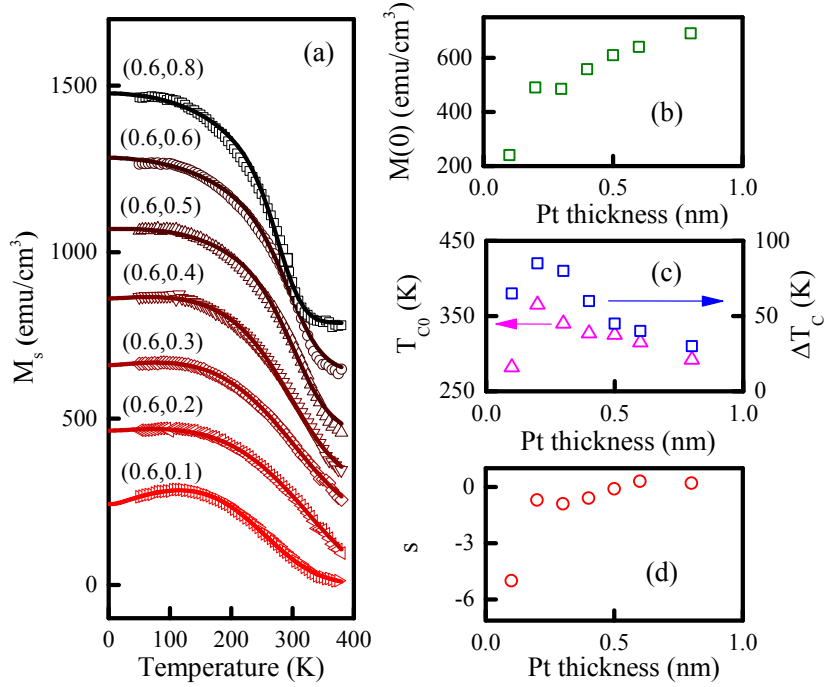

FIG. 4. (a) Experimental *M-T* curves (open symbols) and fitted results (solid lines). The experimental data are the same as those shown in Fig. 3b, but are shifted for clarity (except for the $t_{Pt}$ = 0.1 nm sample). (b) $M_0$, (c) $T_{C0}$ (triangle) and $\Delta T_C$ (square), and (d) *s*, as a function of $t_{Pt}$ obtained from the fittings.

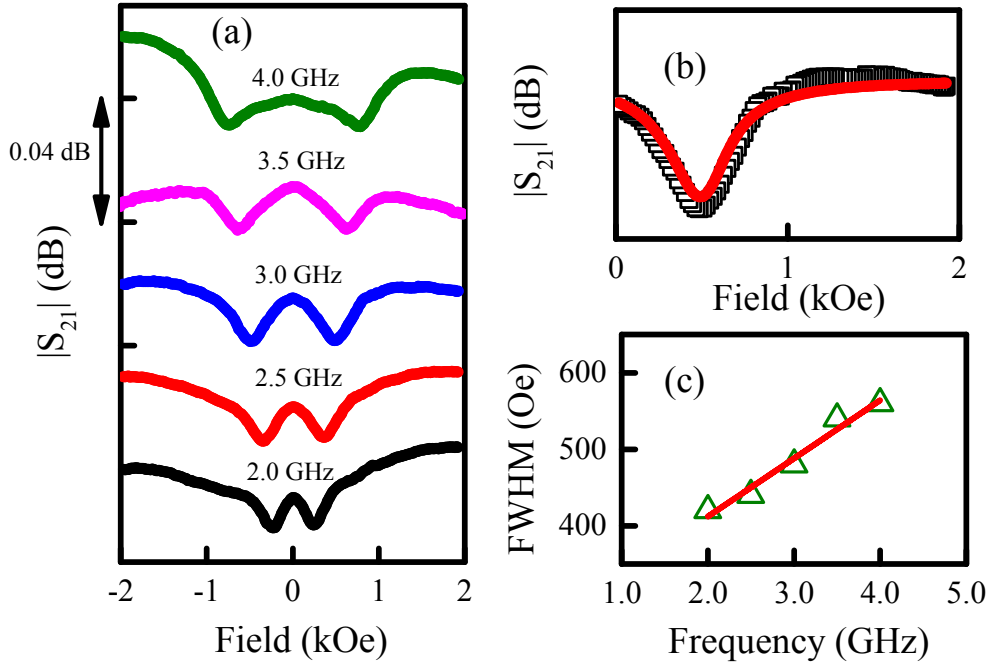

FIG. 5. (a) FMR spectra of Pt(1)/[FeMn(0.6)/Pt(0.5)]$_{80}$ at fixed frequency ranging from 2 GHz to 4 GHz. (b) Data (square symbol) and fitting (line) for FMR signal at $f$ = 3 GHz. (c) Full width at half maximum of the resonance peak (triangle symbol) are plotted againt the frequency. The solid line is a linear fit to the data.

**C. FMR measurements and damping constant**

Magnetic damping plays a key role in the magnetization dynamics of magnetic materials, which can be treated phenomenologically by including a damping term $\alpha \vec{M} \times (d\vec{M}/dt)$ in the Landau-Lifshitz-Gilbert (LLG) equation. Here, α is the Gilbert damping constant, which characterizes the strength of damping. It is commonly assumed that the origin of Gilbert damping is spin-orbit coupling (SOC), same as that of magnetic anisotropy. Since SOC is also the origin of spin-orbit torque, naturally it would be of interest to measure the damping constant of FeMn/Pt multilayers and correlate it with SOT or ISHE. The effective damping constant, including both intrinsic and extrinsic contributions, can be deduced from the FMR line width as a function of resonance frequency. Figure 5a shows the FMR spectra of a Pt(1)/[FeMn(0.6)/Pt(0.5)]$_{80}$ multilayer extracted by VNA at different frequencies ranging from 2 GHz to 4 GHz with a sweeping DC magnetic field. Compared with a homogeneous FM layer, the FMR peak is rather broad. This is presumably caused by the variation in $T_C$ and $M_s$ throughout the multilayer as discussed above. Nevertheless, the average resonance fields at different frequencies can still be described by the Kittel equation[20]

$$2\pi f = \mu_0 \gamma \sqrt{H_{FMR}(H_{FMR} + M_s)} \qquad (8)$$

where $f$ is the frequency, $\gamma$ is the effective gyromagnetic ratio, $M_s$ is the saturation magnetization, $H_{FMR}$ is the resonance field, and $\mu_0$ is the vacuum permeability. The FMR spectra near the resonance region can be roughly fitted by the superposition of a symmetric and an antisymmetric peak. As an example, Fig. 5b shows the fitting result at $f$ = 3 GHz for Pt(1)/[FeMn(0.6)/Pt(0.5)]$_{80}$. The full width at half maximum (FWHM) of the symmetric peak with Lorentz shape is plotted in Fig. 5c (empty square) as a function of frequency. The solid-line is the linear fitting to the relation[21]

$$\Delta H(f) = (\frac{4\pi}{\mu_0 \gamma})\alpha f + \Delta H_0 \qquad (9)$$

where $\alpha$ is the effective damping parameter and $\Delta H_0$ is zero-frequency linewidth caused by magnetic inhomogeneity of the sample. The large $\Delta H_0$ value is consistent with distribution of $T_C$ discussed in IIIB. From the linear fitting, we obtained an effective damping parameter of 0.106 for this specific sample, which is around one order of magnitude larger than that of permalloy at the same thickness,[22] but is comparable to that of Pt/Co multilayers.[23,24] This affirms our previous argument of the twofold role of Pt in Pt/FeMn multilayers[7], *i.e.*, it promotes global FM ordering via proximity effects at Pt/FeMn interfaces and at the same time it functions simultaneously as both a spin current generator and an absorber. It is postulated that both the proximity effect and spin-current absorption contribute to the enhancement of $\alpha$,[25,26] though it is difficult to determine which factor is dominant. When Pt is magnetized, it will be a FM with large SOC which will lead to large damping. FeMn is known to have a small SOC. However, being sandwiched by Pt in the multilayer structure, the precession of its magnetization under ferromagnetic resonance will pump spin current into the neighboring Pt layers, which again will lead to the enhancement of damping. Although a large damping constant is undesirable for applications which require the use of spin torque transferred from other layers to switch its magnetization, it can be effectively exploited for SOT-based applications, *i.e.*, to generate SOT internally by a charge current. This is exactly what we have reported in our earlier work, in which we have demonstrated that it is possible to switch the magnetization of FeMn/Pt multilayers by SOT without any external field.[7] It is worth pointing out that the damping parameter extracted above may be overestimated considering the fact that sample inhomogeneity may also contribute to the large FM linewidth.

**D. Inverse spin Hall effect**

In the aforementioned FMR measurements, we attribute the enhancement of $\alpha$ partially to the absorption of spin current by the Pt layers. As we will discuss shortly in the SMR experiments, for multilayers with relatively thick Pt and FeMn, we may treat them as consisting of alternating FM and HM layers. However, if the Pt and FeMn layers are ultrathin, it is more appropriate to treat the multilayer equivalently as a single FM layer. We consider the multilayer case first. If we focus on a specific FeMn layer inside the multilayer structure, there are two interfaces with the adjacent Pt layers. To differentiate these two interfaces, we call Pt/FeMn the upper interface and FeMn/Pt the lower interface. These two interfaces are not necessarily to be identical due to the large lattice mismatch between Pt and FeMn.[27] Although the FeMn/Pt multilayer behaves like a single phase FM, the magnetic moment is presumably mainly from the FeMn layer. Under the FMR condition, the precession of magnetization in the FeMn layer pumps spin current into the adjacent Pt layers, which is subsequently absorbed either completely or partially depending on the Pt layer thickness. This leads to the enhancement of damping constant as discussed above. If the two interfaces are symmetrical, there should not be a net spin current following inside the multilayer after we take into account the contributions of all the individual layers. However, if the two interfaces are asymmetrical and have different spin-mixing conductance, a net spin current will be generated due to broken inversion symmetry. When this happens, a transverse electromotive force (EMF) will be generated due to ISHE, which can be detected as a voltage signal under open circuit condition. In this context, we have measured the voltage across the two side-contacts of the sample simultaneously with the FMR measurements.

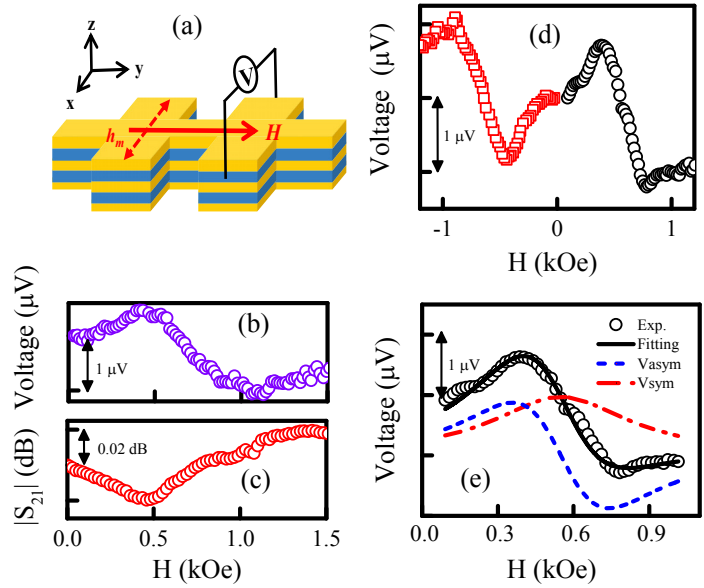

FIG. 6. (a) Measurement geometry of ISHE and FMR. (b) ISHE and (c) FMR spectra for Pt(1)/[FeMn(0.6)/Pt(0.5)]$_{50}$ measured at 3.0 GHz. (d) Voltage signal as a function of positive (circle) and negative (square) magnetic field for Pt(1)/[FeMn(0.6)/Pt(0.4)]$_{10}$ at 3 GHz. (e) Decomposition of measured voltage signal for Pt(1)/[FeMn(0.6)/Pt(0.4)]$_{10}$ at 3GHz into symmetric and antisymmetric components. Symbols are raw data as shown in (d). Dash dotted and dashed lines show the symmetric and antisymmetric components, respectively. The solid-line shows the combined fitting results.

Fig. 6a shows the measurement geometry, where $h_m$ is the *rf* driving field and $H$ is the external field. The measurement was firstly performed on multilayer sample Pt(1)/[FeMn(0.6)/Pt(0.5)]$_{50}$ with $n = 50$. The peak position of the ISHE signal in Fig. 6b and FMR spectrum in Fig. 6c show a good correspondence with each other, suggesting that the ISHE signal might be directly related to FMR absorption. Following that, we carried out the same measurements on Pt(1)/[FeMn(0.6)/Pt($t_{Pt}$)]$_{10}$ samples with $t_{Pt}$ ranging from 0.1 nm to 0.8 nm, respectively. Although the FMR signal of the sample with $n = 10$ was too weak to be detected due to small absorption, the voltage could still be detected for samples with relatively large $M_s$ at RT with $t_{Pt} = 0.2 – 0.5$ nm; however, we could not detect any voltage signal for samples with $t_{Pt} = 0.1$ nm, 0.6 nm and 0.8 nm due to the small $M_s$ at RT. As an example, Fig. 6d shows the measured voltage for Pt(1)/[FeMn(0.6)/Pt(0.4)]$_{10}$ as a function of external magnetic field at fixed frequency of 3 GHz. As can be seen, the peak contains both symmetric and antisymmetric components with respect to the resonance field and its polarity changes when field reverses. Although the transverse voltage can be readily detected under FMR, analysis of the signal is not straightforward because, in addition to ISHE, it also contains contributions due to non-ISHE related effects such as spin rectification effect (SRE) and anomalous Nernst effect (ANE). The ANE is caused by the temperature gradient due to microwave heating and, as reported in several studies, is generally smaller than the SRE effect.[28,29] The SRE signal contains both anisotropic magnetoresistance (AMR) and anomalous Hall effect (AHE) contributions and exhibits complex symmetry and sign dependence on the applied external field, $H$. Based on previous FMR studies in different measurement geometries,[30,31] there are mainly three contributions to the measured voltage signal in the present case: (i) symmetric component due to the ISHE, (ii) symmetric component due to AHE, and (iii) antisymmetric component due to AMR. Based on this, we firstly decompose the obtained voltage signal into the symmetric and antisymmetric components. Fig. 6e shows the symmetric and antisymmetric voltage components of the sample Pt(1)/[FeMn(0.6)/Pt(0.4)]$_{10}$. In this specific case, the peak value of the symmetric component is around 0.97 µV. Based on its symmetry and polarity, the symmetric component should contain both ISHE and AHE contributions. As our experimental setup does not allow us to perform accurate angle-dependent measurement, here we estimate the magnitude of AHE signal using known parameters. Following Chen et al.,[32] the Lorentzian contribution of AHE is approximately given by

$$V_{AHE,L} = \frac{I_{rf,s} \Delta R_{AHE} h_m \cos\theta_0 \cos\Phi}{2\alpha(2H_{FMR} + M_s)} \qquad (10)$$

where $I_{rf,s} = I_{rf,0} R_{wg}/R_s$ with $I_{rf,0}$ the magnitude of rf driving current and $R_{wg}$, $R_s$ the resistance of coplanar waveguide and sample, respectively, $\Delta R_{AHE}$ is the anomalous Hall resistance, $M_s$ is the saturation magnetization, $H_{FMR}$ is the resonant magnetic field, $h_m$ is the rf magnetic field along x direction, $\theta_0$ is the angle between the direction of external magnetic field and coplanar waveguide, and $\Phi$ is the phase of rf field with respect to rf driving current. In the present case, $I_{rf,s} \approx 0.23 \pm 0.03$ mA (calculated from the microwave power assuming maximum delivery efficiency), $R_{AHE} \approx 1.06 \pm 0.11$ Ω (from static measurement), $M_s \approx 262.4 \pm 2.8$ emu/cm$^3$, $H_{FMR} \approx 548.7 \pm 9.2$ Oe, $h_m \approx 36.8 \pm 4.7$ Oe (calculated from rf current), $\theta_0 \approx 0°$ and $\alpha \approx 0.106 \pm 0.01$. Based on these parameters, we obtain $V_{AHE,L} \approx (1.79 \pm 0.8) \times 10^{-7}$ V, which is around one order of magnitude smaller than the measured symmetric voltage component. Since the phase difference between the rf field and rf current is unknown, we assume $\Phi = 0$ in the calculation, which might have led to a slight overestimation of the AHE signal. Based on the discussion above, we believe that the symmetric component of the measured voltage signal is mainly from the ISHE. Before we end this section, it is worth pointing out that the above discussion based on asymmetry in upper and lower interfaces may not apply to multilayers with ultrathin FeMn and Pt as the interfaces are not well defined. This poses a question as to whether the ISHE signal can still be detected in these kinds of samples. As we will discuss in the SMR section, we believe that in this case, we still can detect the ISHE due to extrinsic spin Hall and inverse spin Hall effect.

## E. SMR measurement

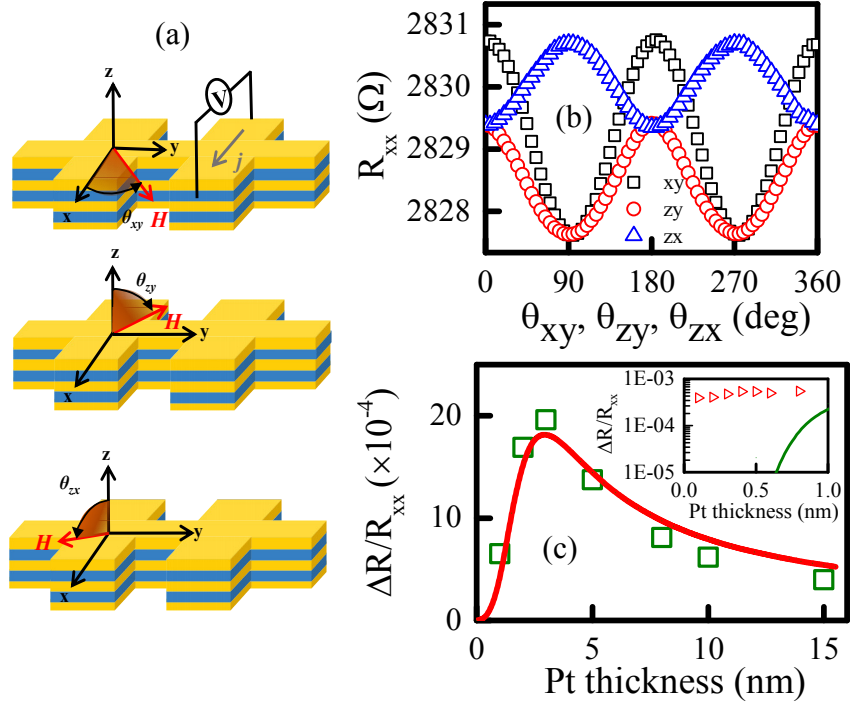

FIG. 7. (a) Geometry of angle-dependent MR measurement. (b) Angle-dependent MR of Pt(1)/[FeMn(0.6)/Pt(0.3)]$_{10}$. (c) Data (square symbol) and fitting (line) of SMR ratio as a function of $t_{Pt}$ for FeMn(3)/Pt($t_{Pt}$) bilayers. Inset shows the calculated SMR (line) for FeMn(0.6)/Pt($t_{Pt}$) bilayers at small Pt thickness as well as experimentally obtained SMR ratio (triangle symbol) for Pt(1)/[FeMn(0.6)/Pt($t_{Pt}$)]$_{10}$ multilayers.

Both FMR and ISHE measurements confirm that spin current generation and absorption occur simultaneously in the multilayer. This is exactly the ingredient for generating both SOT and SMR, which themselves are complementary processes of each other.[33] To confirm this, we have performed SMR measurements for the same batch of samples used for the ISHE measurements. Figure 7a shows the geometry of the SMR, or angle-dependent magnetoresistance (MR) measurements, which were carried out with an applied field of 30 kOe rotating in the *zy*, *zx*, and *xy* planes, respectively. All the multilayer samples exhibit clear SMR signal. As a typical example, Figure 7b shows the angle-dependent MR of Pt(1)/[FeMn(0.6)/Pt(0.3)]$_{10}$. From the angle-dependence, we can see that only AMR is observed when the field is rotated in the *zx* plane, whereas the signal obtained in the *zy* plane is dominantly from SMR. When the field is rotated in the *xy* plane, both AMR and SMR are detected. Recently, Manchon developed a model for SMR in AFM/HM bilayer,[34] which applies to the collinear AFM with well-defined Neel order $\vec{n} = \vec{m}_1 - \vec{m}_2$, where $\vec{m}_1$, $\vec{m}_2$ are the unit vector of the two spin sublattices, respectively. According to this model, the SMR of AFM/HM bilayers is given by

$$\frac{\Delta R}{R_{xx}} = \frac{\lambda_N \sigma_N}{d_N \sigma_N + d_{AF} \sigma_{AF}} \theta_{SH}^2 (1 - \cosh^{-1} \frac{d_N}{\lambda_N})^2 \frac{(\gamma_\| \eta_\| - \gamma_\perp \eta_\perp)}{(1 + \gamma_\| \eta_\| \tanh^{-1} \frac{d_N}{\lambda_N})(1 + \gamma_\perp \eta_\perp \tanh^{-1} \frac{d_N}{\lambda_N})} \quad (11)$$

with $\eta_{\|,\perp} = 1 + (r_{\|,\perp} \sigma_{\|,\perp}^{AF} / \lambda_{\|,\perp}^{AF}) \tanh(d_{AF}/\lambda_{\|,\perp}^{AF})$, $\gamma_{\|,\perp} = (\lambda_{\|,\perp}^{AF} \sigma_N / \lambda_N \sigma_{\|,\perp}^{AF}) \tanh^{-1}(d_{AF}/\lambda_{\|,\perp}^{AF})$, $\lambda_\|^{AF} = \sqrt{D_\|^{AF} \tau_{sf}^{AF}}$ and $\lambda_\perp^{AF} = \sqrt{D_\perp^{AF} (1/\tau_{sf}^{AF} + 1/\tau_\varphi^{AF})}$. Here, the subscript $\|(\perp)$ refers to the configuration when the spin polarization aligns parallel (transverse) to the Neel order parameter, $\theta_{SH}$ is the spin Hall angle, $D^{AF}$ is the electron diffusion coefficient in the AFM, $\tau_{sf}^{AF}$ is the conventional isotropic spin relaxation time, $\tau_\varphi^{AF}$ is the spin dephasing time that relaxes only the spin component that is transverse to the Neel order parameter, *r* is the interfacial resistivity, $\lambda_N$, $\sigma_N$ ($\sigma_{AF}$) and $d_N$ ($d_{AF}$) are spin diffusion length, conductivity and thickness of the HM (AFM) layer, respectively. As shown in Fig. 7c, by varying the Pt thickness systematically, we found that the thickness-dependence of SMR of FeMn(3)/Pt($t_{Pt}$) bilayers can be fitted well using the following parameters: $\lambda_N$ = 1.05 ± 0.05 nm, $\lambda_\|^{FeMn}$ = 4.1 ± 0.1 nm, $\lambda_\perp^{FeMn}$ = 1.70 ± 0.07 nm, $\theta_{SH}$ = 0.28 ± 0.03, $\tau_{sf}^{FeMn}$ = (4.25 ± 0.25) × 10$^{-14}$ s, $\tau_\varphi^{FeMn}$ = (7.75 ± 0.25) × 10$^{-15}$ s, $\sigma_N$ = 4.0 × 10$^6$ S/m, $\sigma_\|^{FeMn}$ = 1.0 × 10$^6$ S/m, and $\sigma_\perp^{FeMn}$ = 1.5 × 10$^6$ S/m. The inset of Fig. 7c shows the region with small Pt thickness in log-scale, together with the experimental SMR of Pt(1)/[FeMn(0.6)/Pt($t_{Pt}$)]$_{10}$ multilayers. As can be seen, the experimental SMR values for multilayers are significantly larger than the simulated results for bilayers, particularly at very small Pt thickness. The difference becomes smaller when the Pt thickness increases. This suggests that when Pt is thick, the multilayer can be considered as comprising of magnetically

decoupled bilayers and therefore the SMR ratio should be the same for both types of samples. However, when $t_{Pt}$ is very small, the multilayer behaves more like a "single phase" FM; this is the reason why the SMR is different from that of bilayers with small $t_{Pt}$. The observation of large SMR in the multilayers suggests that there is spin current generation/absorption process taking place inside the multilayer, presumably due to either intrinsic (for samples with thick Pt) or extrinsic SHE/ISHE (for samples with ultrathin Pt) or the combination of both. This is also the reason why a large SOT was observed in these structures.

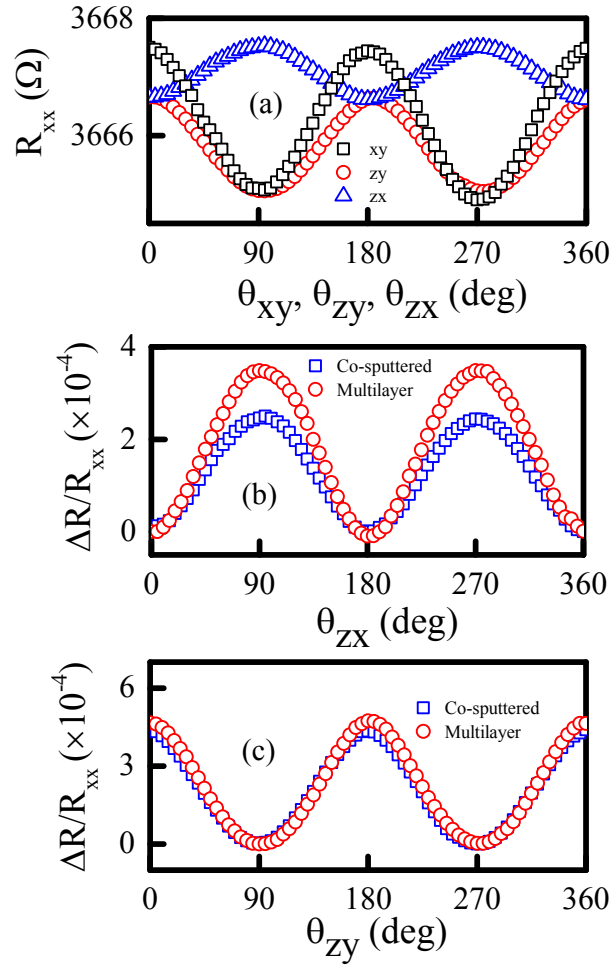

FIG. 8. (a) Angle-dependent MR of co-sputtered sample; (b) AMR and (c) SMR of co-sputtered and multilayer samples with same nominal composition and thicknesses.

To shed some light on the origin of SMR, particularly, in structures with ultrathin Pt layers, we have also fabricated and measured the SMR of co-sputtered samples. At the same nominal thickness and composition, the co-sputtered sample is more resistive than its multilayer counterpart, consistent with its more disordered structure. Despite the structural difference, SMR of similar magnitude of that of multilayers was also observed in co-sputtered samples. Fig.8 (a) shows the angle-dependent MR of a co-sputtered FeMn:Pt sample with overall nominal thickness of $t_{FeMn}$ = 6 nm and $t_{Pt}$ = 3 nm (calculated from the deposition power and duration). Both the AMR and SMR components are present in the angle-dependent MR. In Figs. 8b and 8c, we show the normalized AMR and SMR curve for both the co-sputtered and Pt(1)/[FeMn(0.6)/Pt(0.3)]$_{10}$ multilayer sample. The nominal thickness and composition are the same for the two samples and both samples are capped with a 1 nm Pt. We have confirmed that the SMR for Pt(1)/[FeMn(0.6)/Pt(0.3)]$_{10}$ and [FeMn(0.6)/Pt(0.3)]$_{10}$ is almost the same, and therefore, the 1 nm Pt capping layer is not responsible for the SMR observed in both cases. Although further studies are required to elucidate the SMR mechanism in both co-sputtered and multilayer samples with ultrathin layers, the observed SMR can be qualitatively explained using the drift-diffusion model by taking into account both the precession and dephasing of SHE-generated spin inside a single FM with large spin-orbit coupling. The dynamics of the SHE-generated spin accumulation $\hat{S}$ is governed by the coupled equations:[35]

$$\nabla^2 \hat{S} = \frac{1}{\lambda_\varphi^2} \hat{S} \times \hat{m} + \frac{1}{\lambda_\theta^2} \hat{m} \times (\hat{S} \times \hat{m}) + \frac{1}{\lambda_S^2} \hat{S} \quad (12a)$$

$$\hat{J}_i^S = -D(\nabla \hat{S}_i + \theta_{SH} \hat{e}_i \times \nabla n) \quad (12b)$$

where $\hat{S}$ is the non-equilibrium spin density generated by SHE, $\hat{m}$ is direction of the local magnetization, $\lambda_\varphi$, $\lambda_\theta$ and $\lambda_S$ are spin precession, dephasing and spin-flip diffusion length, respectively, $\hat{J}_i^S$ is the $i^{th}$ component of spin current with polarization in $\hat{S}$ direction, $D$ is the diffusion coefficient, $n$ is the charge density, $\theta_{SH}$ is the spin Hall angle, and $\hat{e}_i$ is a unit vector. The angle-dependence of MR (or simply SMR) appears due to the additional electromotive force generated by $\hat{J}_i^S$ via ISHE. Eq. 12a can be better understood by considering the special cases: i) $\lambda_\varphi, \lambda_\theta \gg \lambda_S$, ii) $\lambda_\theta \gg \lambda_S, \lambda_\varphi$, and iii) $\lambda_S \gg \lambda_\theta, \lambda_\varphi$. In case i), the first two terms at the right-hand-side of Eq. 12a can be ignored, which leads to the spin diffusion equation for a non-magnetic metal. In this case, there will be no SMR-like angle-dependent MR unless when it is in contact with a ferromagnetic layer. In case ii), the 2$^{nd}$ term can be ignored, which leads to $\nabla^2 \hat{S} = \frac{1}{\lambda_\varphi^2} \hat{S} \times \hat{m} + \frac{1}{\lambda_S^2} \hat{S}$. This is similar to the case of Hanle MR (HMR) in HM except that the spin precession in HMR is caused by an external field,[36] whereas in the present case it is caused by the exchange field of the FM itself. In the last case, both spin precession and dephasing terms have to be taken into account on equal footing. To estimate the influence of these two terms on the spin density, we consider two special cases which is related to the transverse and vertical MR, *i.e.,* i) $\hat{m} = (0,1,0)$ and ii) $\hat{m} = (0,0,1)$. In the thin film geometry, we are mainly concerned about the spin accumulation on the top and bottom surfaces which have a spin polarization dominantly in *y*-direction. In this case, when $\hat{m} = (0,1,0)$, both the precession and dephasing terms can be ignored. Under this condition, spin accumulation occurs on both surfaces, resulting in a diffusion spin current reflected back to the sample. This will lead a smaller resistivity due to ISHE effect. On the other hand, when $\hat{m} = (0,0,1)$, the dephasing and diffusion term can be combined, leading to $\nabla^2 \hat{S} = \frac{1}{\lambda_\varphi^2} \hat{S} \times \hat{m} + \frac{1}{\lambda^2} \hat{S}$, where $\frac{1}{\lambda^2} = \frac{1}{\lambda_\theta^2} + \frac{1}{\lambda_S^2}$. This expression is similar to the case of HMR except that the spin diffusion length is replaced by an equivalent diffusion length. We can let $\lambda_S = \lambda_\parallel^{FeMn} = 4.1$ nm and $\lambda_\theta = \lambda_\varphi = \lambda_\perp^{FeMn} = 1.7$ nm, and then $\lambda = 1.57$ nm. Since this equation is similar to the case of HMR, we can use the solution given in the supplementary material of S. Vélez *et al.*[36] to estimate the SMR-like resistance change due to the first term, which is given by

$$\frac{\Delta R}{R_{xx}} \approx \frac{3\theta_{SH}^2}{4} (\frac{\lambda}{\lambda_\varphi})^2 \frac{\lambda}{d} \tag{13}$$

where $d$ is the sample thickness, $\Delta R$ is the change in longitudinal resistance and $R_{xx}$ is the longitudinal resistance at zero field. Using $\theta_{SH} = 0.1$, $d = 10$ nm, $\lambda_\varphi = 1.7$ nm and $\lambda = 1.57$ nm, we obtain an MR ratio $\frac{\Delta R}{R_{xx}} = 0.1\%$, which is on the same order of magnitude of SMR observed experimentally. Although the exact value depends on the parameters used, we believe it does explain the salient feature of the MR response observed in both the co-sputter and multilayer samples with ultrathin Pt and FeMn layer. However, when the Pt layer is sufficiently thick, the bilayer model seems to be more appropriate as manifested in the agreement between experiment and theoretical model shown in Fig.7.

**F. Discussion**

In this study, we investigated the static and dynamic properties of [FeMn/Pt]$_n$ multilayers by combined techniques of magnetometry, FMR, ISHE and SMR, and found a good correlation in the results obtained by different techniques. First, the FMR and ISHE signals can only be detected in samples with sufficiently large $M_s$ at room temperature, which typically happens in samples with a large repetition period, and magnetic inhomogeneity due to thickness-sensitive $T_c$ variation is well reflected in the broad peak appeared in the FMR and ISHE spectra. Second, the FMR peak positions correspond well with those of ISHE. Third, SMR with a magnitude comparable to that of FeMn/Pt bilayer was observed, supporting the presence of large SOT. All these results in combination with the fact that the multilayer behaves like a 3D Heisenberg ferromagnet and exhibits a large SOT seem to suggest that there is a broken inversion symmetry (BIS) inside the multilayers. The most likely origin of the BIS in the multilayer is the crystalline asymmetry of the FeMn/Pt and Pt/FeMn interface caused by the different atomic size. According to Liu *et al.*,[27] the atom radii of Pt and FeMn are 0.139 nm and 0.127 nm, respectively. When depositing Pt on fcc (111) textured FeMn layer, the crystal direction and atom packing will have to change in order to accommodate the large Pt atoms as the (111) plane is already close-packed. On the other hand, the situation will be different when smaller Fe and Mn atoms are deposited on fcc (111) textured Pt layer. This will lead to local inversion asymmetry in the multilayer. Similar phenomenon has also been reported for Co/Pt[37,38] and Co/Pd[39] multilayers. This explains why a large SOT is generated when a charge current is applied to the multilayer, as we demonstrated previously. However, the observation of SMR in co-sputtered samples with a magnitude comparable to the multilayer suggests the observed phenomena can also be explained by simultaneous actions of extrinsic SHE and ISHE, particularly in multilayers with ultrathin FeMn and Pt. Further studies are required to evaluate the relative contribution of intrinsic and extrinsic SHE and ISHE in FeMn/Pt multilayers with different thickness combinations.

## IV.   CONCLUSIONS

The static and dynamic magnetic properties of FeMn/Pt multilayers have been studied using combined techniques of magnetometry, FMR, ISHE and SMR. Despite the fact that FeMn is an AFM in the bulk phase, FeMn/Pt multilayers with ultrathin FeMn ($t_{FeMn}$ < 0.8 nm) and Pt ($t_{Pt}$ < 1.0 nm) layers exhibit ferromagnetic properties with in-plane magnetic anisotropy. The temperature dependence of saturation magnetization can be fitted well using a phenomenological model developed for 3D Heisenberg magnet by including a finite distribution in $T_C$. The latter is attributed to the high sensitivity of magnetic properties to subtle changes in the individual layer thicknesses. The finite distribution of $T_C$ correlates well with the broad absorption peaks observed in the FMR spectra. A large damping parameter (~ 0.106) is derived from the frequency dependence of FMR linewidth, which is comparable to the values reported for Co/Pt multilayers. Clear ISHE signals and SMR have been observed in all samples below the Curie temperature, which corroborate the strong SOT effect observed previously. The latter is attributed to the crystalline asymmetry between the top FeMn/Pt and bottom Pt/FeMn interfaces when the Pt layer is relatively thick. However, for samples with ultrathin Pt, extrinsic SHE/ISHE may play a more important role in the phenomena observed.

**ACKNOWLEDGMENTS**

Y.H.W. would like to acknowledge support by the Singapore National Research Foundation, Prime Minister's Office, under its Competitive Research Programme (Grant No. NRF-CRP10-2012-03) and Ministry of Education, Singapore under its Tier 2 Grant (Grant No. MOE2013-T2-2-096). Y.H.W. and J.S.C. are members of the Singapore Spintronics Consortium (SG-SPIN).

**FIGURE CAPTIONS**

FIG. 1. X-ray diffraction pattern of Pt(1)/[FeMn(0.6)/Pt(0.4)]$_{10}$. Dotted lines indicate the (111) peak position of Pt and FeMn, respectively.

FIG. 2. XRR patterns of Pt(1)/[FeMn(0.6)/Pt(0.6)]$_{30}$ multilayer sample (red solid-line) and co-sputtered sample (blue dotted-line) deposited under the same condition.

FIG. 3. (a) Hysteresis loop of Pt(1)/[FeMn(0.6)/Pt(0.3)]$_{10}$ measured at 50 K (square) and 300 K (circle), respectively. (b) Saturation magnetization as a function of temperature. The legend ($t_1,t_2$) denotes a multilayer with a FeMn thickness of $t_1$ and Pt thickness of $t_2$. The number of period for all samples is fixed at 10.

FIG. 4. (a) Experimental *M-T* curves (open symbols) and fitted results (solid lines). The experimental data are the same as those shown in Fig. 3b, but are shifted for clarity (except for the $t_{Pt}$ = 0.1 nm sample). (b) M$_0$, (c) T$_{C0}$ (triangle) and ΔT$_C$ (square), and (d) *s,* as a function of $t_{Pt}$ obtained from the fittings.

FIG. 5. (a) FMR spectra of Pt(1)/[FeMn(0.6)/Pt(0.5)]$_{80}$ at fixed frequency ranging from 2 GHz to 4 GHz. (b) Data (square symbol) and fitting (line) for FMR signal at *f* = 3 GHz. (c) Full width at half maximum of the resonance peak (triangle symbol) are plotted versus the frequency. The solid line is a linear fit to the data.

FIG. 6. (a) Measurement geometry of ISHE and FMR. (b) ISHE and (c) FMR spectra for Pt(1)/[FeMn(0.6)/Pt(0.5)]$_{50}$ measured at 3.0 GHz. (d) Voltage signal as a function of positive (circle) and negative (square) magnetic field for Pt(1)/[FeMn(0.6)/Pt(0.4)]$_{10}$ at 3 GHz. (e) Decomposition of measured voltage signal for Pt(1)/[FeMn(0.6)/Pt(0.4)]$_{10}$ at 3GHz into symmetric and antisymmetric components. Symbols are raw data as shown in (d). Dash dotted and dashed lines show the symmetric and antisymmetric components, respectively. The solid-line shows the combined fitting results.

FIG. 7. (a) Geometry of angle-dependent MR measurement. (b) Angle-dependent MR of Pt(1)/[FeMn(0.6)/Pt(0.3)]$_{10}$. (c) Data (square symbol) and fitting (line) of SMR ratio as a function of $t_{Pt}$ for FeMn(3)/Pt($t_{Pt}$) bilayers. Inset shows the calculated SMR (line) for FeMn(0.6)/Pt($t_{Pt}$) bilayers at small Pt thickness as well as experimentally obtained SMR ratio (triangle symbol) for Pt(1)/[FeMn(0.6)/Pt($t_{Pt}$)]$_{10}$ multilayers.

FIG. 8. (a) Angle-dependent MR of co-sputtered sample; (b) AMR and (c) SMR of co-sputtered and multilayer samples with same nominal composition and thicknesses.